\documentclass[twocolumn,superscriptaddress,floatfix,showpacs,prb]{revtex4}
\usepackage{graphicx,amsfonts,amssymb,amsmath,hyperref,hypcap,enumerate}
\usepackage{tikz}

\newcommand{\beq}{\begin{equation}}
\newcommand{\eeq}{\end{equation}}
\newcommand{\bag}{\begin{aligned}}
\newcommand{\eag}{\end{aligned}}
\newcommand{\dt}{{\text d}}
\newcommand{\stxt}{{\text s}}
\newcommand{\vtxt}{{\text v}}
\newcommand{\Ctxt}{{\text C}}

\newcommand{\Ztxt}{{\text Z}}
\newcommand{\Bc}{\mathcal{B}}
\newcommand{\Hc}{\mathcal{H}}
\newcommand{\Ec}{\mathcal{E}}
\newcommand{\Lc}{\mathcal{L}}
\newcommand{\al}{\alpha}
\newcommand{\rb}{{\bf r}}
\newcommand{\dg}{\dagger}
\newcommand{\pd}{\partial}
\newcommand{\ga}{\gamma}
\newcommand{\Tr}{\text{Tr}}

\begin{document}
\title{SO(5) symmetry in the quantum Hall effect in graphene}

\author{Fengcheng Wu}
\affiliation{Department of Physics, University of Texas at Austin, Austin, TX 78712, USA}

\author{Inti Sodemann\footnote{Current address: Department of Physics, Massachusetts Institute of Technology, Cambridge, MA 02139, USA}}
\affiliation{Department of Physics, University of Texas at Austin, Austin, TX 78712, USA}

\author{Yasufumi Araki}
\affiliation{Department of Physics, University of Texas at Austin, Austin, TX 78712, USA}

\author{Allan H. MacDonald}
\email{macdpc@physics.utexas.edu}
\affiliation{Department of Physics, University of Texas at Austin, Austin, TX 78712, USA}

\author{Thierry Jolicoeur}
\email{thierry.jolicoeur@u-psud.fr}
\affiliation{Laboratoire de Physique Th\'eorique et Mod\`eles statistiques,CNRS and Universit\'e Paris-Sud, Orsay 91405, France }

\begin{abstract}
Electrons in graphene have four flavors associated with low-energy spin and valley degrees of freedom.
The fractional quantum Hall effect in graphene is dominated by long-range Coulomb interactions
which are invariant under rotations in spin-valley space.  This SU(4) symmetry is spontaneously broken
at most filling factors, and also weakly broken
by atomic scale valley-dependent and valley-exchange interactions with coupling
constants $g_{z}$ and $g_{\perp}$.  In this paper we demonstrate that
when $g_{z}=-g_{\perp}$ an exact SO(5) symmetry survives which unifies the
N\'eel spin order parameter of the antiferromagnetic state
and the $XY$ valley order parameter of the Kekul\'e distortion state
into a single five-component order parameter. The proximity of the highly insulating quantum Hall state observed in graphene at $\nu=0$ to an ideal SO(5) symmetric quantum Hall state remains an open experimental question. We illustrate the physics associated with this
SO(5) symmetry by studying the multiplet structure and collective
dynamics of filling factor $\nu=0$ quantum Hall states based on exact-diagonalization and
low-energy effective theory approaches. This allows to illustrate how manifestations of the SO(5) symmetry would survive even when it is weakly broken.
\end{abstract}


\maketitle

\section{Introduction}
Electron-electron interactions in the fractional quantum Hall effect (FQHE) regime give rise to a host of non-perturbative and unexpected
phenomena, including importantly the emergence of quasiparticles with fractional charge and statistics.
In this paper we suggest that neutral graphene in the FQHE regime could also provide a relatively simple
example of the complex many-particle physics that occurs in systems with simultaneous quantum fluctuations of
competing order parameters.  Because each of its Landau levels has a four-fold spin/valley
flavor degeneracy in the absence of Zeeman coupling, large gaps and associated quantum Hall effects
are produced by single-particle physics only at filling factors $\nu = \pm 2, \pm 6, \ldots$.
The quantum Hall effect nevertheless occurs at all intermediate integer filling factors~\cite{Kim06,Young12},
and at many fractional filling factors~\cite{Du09, Bolotin09, Dean11}, usually~\cite{NM} with a broken symmetry
incompressible ground state.
When lattice corrections to the continuum Dirac model's Coulomb interactions are ignored the ground state at neutrality
($\nu=0$) is a Slater determinant\cite{YDM} with all the $N=0$ single-particle states of two arbitrarily chosen flavors occupied
and, because the Hamiltonian is SU(4) invariant, has four independent degenerate Goldstone modes.
The rich flavor physics of graphene in the quantum Hall regime has
already been established by experiments which demonstrate that phase transitions
between distinct many-electron states with the same filling factor $\nu$
can be driven by tuning magnetic field strength or tilt-angle~\cite{Yacoby12, Yacoby13, Pablo14,Halperin13,Inti14}.

In graphene
the competition between states with Kekul\'e-distorion(KD), antiferromagnetic(AF), ferromagnetic(F), charge-density wave(CDW), and other types of
order is controlled by Zeeman coupling to the electron-spin, and also by weak atomic-range
valley-dependent~\cite{BA} interactions.  A variety of approaches have been used to estimate these
short-range corrections to the Coulomb interaction~\cite{Fisher07,Herbut,Lederer07, Jung09, Nomura_KD, Mudry10, Kharitonov_MLG}.
In this paper, we adopt a two-parameter phenomenological
model motivated by crystal momentum conservation and
by the expectation that corrections to the Coulomb interaction are
significant only at distances shorter than a magnetic length~\cite{Kharitonov_MLG}
$l_B=\sqrt{\hbar c/e B_\perp}$.  ($B_\perp$ is the magnetic field component perpendicular to the graphene plane.)
We demonstrate that along a line in this parameter space SU(4) symmetry is reduced
only to a SO(5) subgroup.
In this paper, we take interaction-driven quantum Hall states at $\nu=0$
as an example to illustrate the physical manifestation of the SO(5) symmetry.
We explicitly derive a low-energy theory at $\nu=0$ that is able to
account simultaneously for N\'eel antiferromagnetism and Kekul\'e lattice-distortion order and
demonstrate that along the SO(5) line the four collective modes remain gapless
in spite of the reduced symmetry.
The exact SO(5) symmetry we have identified in graphene's quantum Hall regime
is analogous to the approximate symmetry
conjectured in some models of high-$T_c$ superconductivity~\cite{SO5}.
Our work demonstrates that an enlarged symmetry like SO(5) can indeed be exactly realized in a realistic microscopic Hamiltonian.
In the following, we start with a systematic analysis of Hamiltonian symmetries
and then use both exact-diagonalization and low-energy effective
models at $\nu=0$ to identify some symmetry-related properties.

Although our work focuses on the properties of the quantum Hall state at neutrality, we demonstrate that the SO(5) symmetry is an exact symmetry of the interaction Hamiltonian for the quantum Hall states in the zero energy Landau level of graphene.
Therefore this symmetry is expected to emerge as well in the phase diagrams at arbitrary filling fractions in this Landau level.

The quantum Hall state of graphene at neutrality is believed to be a canted antiferromagnet, as indicated by the behaviour of the edge conductance in experiments with tilted magnetic fields~\cite{Pablo14}. However, as we argue below, these experiments are not sufficient to determine the proximity of graphene to the ideal SO(5) symmetric state. Even if graphene is in the antiferromagnetic side of the phase diagram, the presence of a weakly broken SO(5) symmetry would have important consequences, such as the existence of additional weakly gapped neutral collective modes as we will discuss in detail in Section~\ref{Low-energy} and in Appendix~\ref{Zeeman}.

\section{Hamiltonian symmetries}
When projected to the $N=0$ Landau level (LL)
the graphene Hamiltonian is
\beq
\bag
H\, \, =&\, H_\Ctxt+H_\vtxt+H_\Ztxt,\\
H_\Ctxt=&\, \frac{1}{2}\sum_{i\neq j}\frac{e^2}{\epsilon|\vec{r}_i-\vec{r}_j|},\\
H_\vtxt=&\, \frac{1}{2}\sum_{i\neq j}\big(g_z\tau_z^i\tau_z^j+g_\perp(\tau_x^i\tau_x^j+\tau_y^i\tau_y^j)\big)\delta(\vec{r}_i-\vec{r}_j),\\
H_\Ztxt=&\, -\epsilon_\Ztxt\sum_{i}\sigma_z^i.
\label{Hamiltonian}
\eag
\eeq
In Eq.~(\ref{Hamiltonian}) $H_\Ctxt$ is the valley-independent Coulomb interaction,
$\epsilon$ is an environment-dependent effective dielectric constant,
$H_\vtxt$ is the short-range valley-dependent interaction,
$\tau_\al (\al=x,y,z)$ are Pauli matrices which act in valley space,
$H_\Ztxt$ is the Zeeman energy~\cite{Kim06}, $\epsilon_\Ztxt= \mu_B B$
where  $\mu_B$ is the Bohr magneton and $B$ is the total magnetic field strength,
and $\sigma_\al$ are Pauli matrices which act in spin space.
Note that $B$ can have components both perpendicular and parallel to the graphene plane and that we
have chosen the $\hat{z}$ direction in spin-space to be aligned with $B$.
The form used for $H_\vtxt$ in Eq.~(\ref{Hamiltonian}) was
proposed by Kharitonov~\cite{Kharitonov_MLG,Aleiner07}.

The short-range interaction coupling constants $g_{z,\perp}/l_B^2$
are estimated to be $\sim a_0/l_B$ times the Coulomb energy scale $e^2/\epsilon l_B$,
where $a_0 \sim 0.01 l_B$ is the lattice constant of graphene.
They are therefore weak and physically relevant mainly
when they lift low-energy Coulomb-only model degeneracies.
For later notational convenience we define the energy scales $u_{z,\perp}=g_{z,\perp}/(2\pi l_B^2)$.
The Coulomb interaction $H_\Ctxt$
in Eq.~(\ref{Hamiltonian}) commutes with the fifteen SU(4) transformation
generators which can be chosen as follows~:
\beq
\bag
S_\al&=\frac{1}{2}\sum_i\sigma_\al^i, \; \; \;  \; \; \; T_\al=\frac{1}{2}\sum_i\tau_\al^i,\\
N_\al&=\frac{1}{2}\sum_i\tau_z^i\sigma_\al^i, \;\;\;\Pi_\al^\beta=\frac{1}{2}\sum_i\tau_\beta^i\sigma_\al^i,
\label{generator}
\eag
\eeq
where the indices $\al=x,y,z$ and $\beta=x,y$.
$S_\al$ and $T_\al$ are respectively the total spin and valley pseudospin.
Due to the equivalence between valley and sublattice degrees of freedom
in the $N=0$ LL of graphene, $N_\al$ can be identified as a N\'eel vector.
The physical meaning of the six $\Pi_\al^\beta$ operators
is discussed below.

\begin{figure}[htbp]
\vspace{-0.5cm}
 \includegraphics[width=0.6\columnwidth]{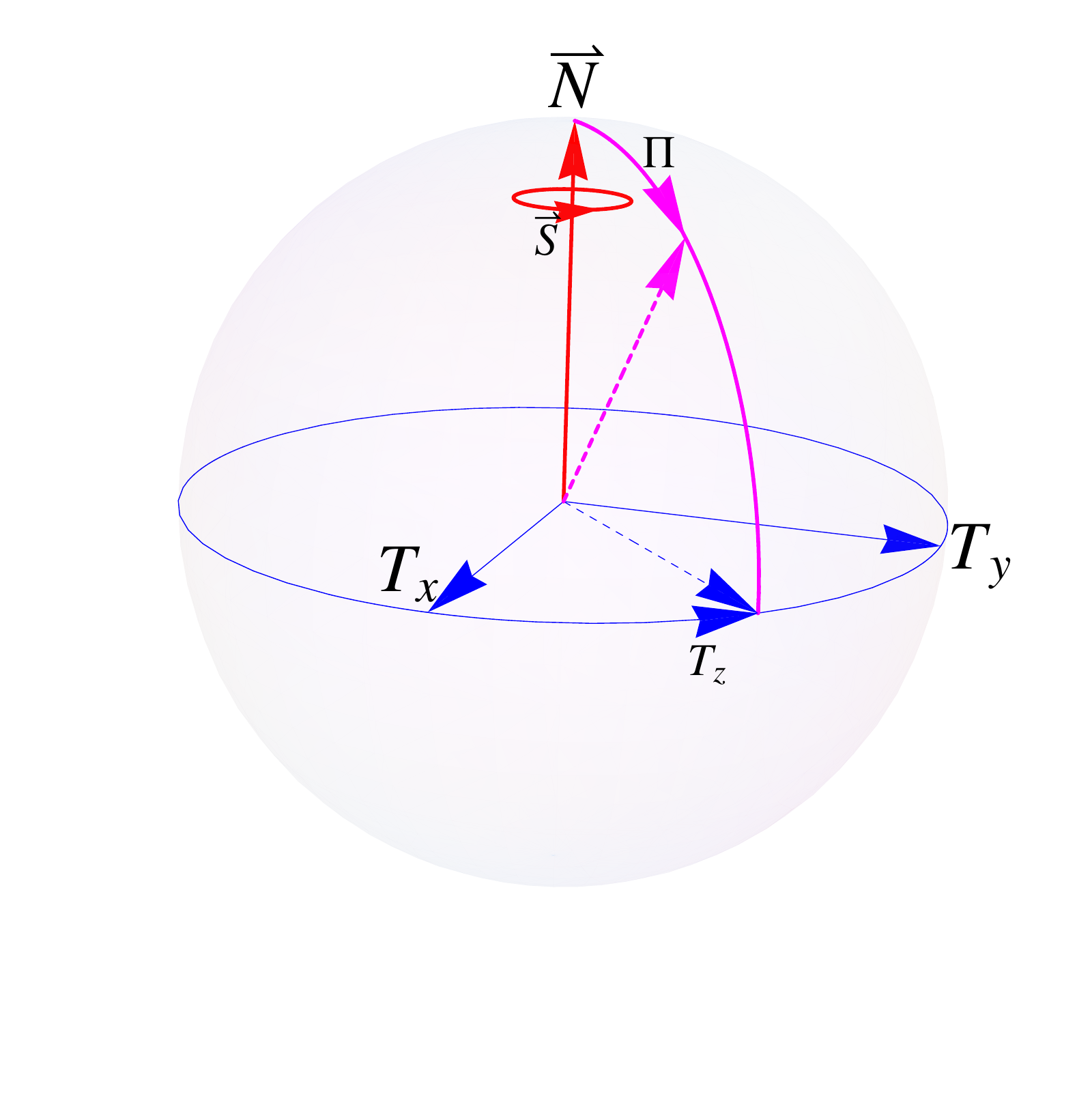}
\vspace{-1.2cm}
 \caption{Schematic illustration of the five component $(T_{x, y}, N_{x, y, z})$
order parameter space, and of rotations in this vector space produced by the SO(5) generators.}
 \label{fig1}
\end{figure}

SU(4) symmetry is broken by the valley-dependent short range interactions.
At a generic point in the ($g_z, g_\perp$) plane, $H_\vtxt$ breaks the SU(4) symmetry
down to SU(2)$_\stxt\times$U(1)$_\vtxt$ with the U(1)$_\vtxt$ symmetry corresponding to
conservation of the valley polarization $T_z$
and the SU(2)$_\stxt$ symmetry corresponding to global spin-rotational invariance.
Two high-symmetry lines in the ($g_z, g_\perp$) parameter space are evident~:
(1) for $g_\perp=0$ the system is invariant under separate spin-rotations in each valley
yielding symmetry group SU(2)$_\stxt^K\times$SU(2)$_\stxt^{K'}\times$U(1)$_\vtxt$ and
(2) for $g_\perp=g_z$ there is a full rotational symmetry in valley space yielding
symmetry group SU(2)$_\stxt\times$SU(2)$_\vtxt$.
We have discovered that there is even higher symmetry along the $g_\perp=-g_z$ line
where the generic SU(2)$_\stxt\times$U(1)$_\vtxt$ symmetry is
enlarged to SO(5)~: see Appendix A for an explicit proof.
Along this line the Hamiltonian commutes with ten ($\vec{S}$, $T_z$, and the six $\Pi$ operators) of the
fifteen SU(4) generators identified in Eq.~(\ref{generator}).
The other five $(T_{x, y}, N_{x, y, z})$ SU(4) generators form a
a natural order-parameter vector space on which the SO(5) group acts.
As illustrated schematically in Fig.~\ref{fig1}, spin operators $\vec{S}$ generate rotations
in the N\'eel vector space $\vec{N}$, $T_z$ generates rotations in the valley $XY$ vector space $T_{x,y}$,
and the $\Pi$ operators generate rotations that connect these two spaces.
When the Zeeman term is added to the Hamiltonian the spin-symmetry is limited
to invariance under rotations about the direction of the magnetic field.
The symmetry groups of $H_\Ctxt+H_\vtxt$ and $H$ and the corresponding generators are listed in Table \ref{table1}.

As we will demonstrate, the SO(5) symmetry is spontaneously broken when it is exact. Provided that the Zeeman
and short-range interaction terms which explicitly breaks SO(5) symmetry is not too strong, the
$(T_{x, y}, N_{x, y, z})$ vectors can be used to construct a useful
Ginzburg-Landau model or quantum effective-field theory.
The N\'eel vector components of the order parameter characterize the AF
part of the order, while the valley $XY$ components capture the KD~\cite{Kharitonov_MLG,Nomura_KD}
part of the order.  The SO(5) symmetry demonstrates that states
which appear quite different at a first glance are close in energy
and that they can be continuously transformed into one another by appropriate rotations in the SO(5)
order parameter space.
The 5D vector $(T_{x, y}, N_{x, y, z})$ identified here
provides a concrete example for the 56 possible quintuplets
proposed in graphene~\cite{Five_2009, Five_2012}.
Although we focus here mainly on monolayer graphene, a similar symmetry analysis applies to the
$N=0$ LL in bilayer graphene~\cite{Kharitonov_BLG,Kharitonov_edge, Kharitonov_AFM}.

\begin{table*}[t]
\caption{Expanded symmetries along high-symmetry lines in the ($g_z$, $g_\perp$) plane.
At a generic point in the ($g_z$, $g_\perp$) plane $H_\Ctxt+H_\vtxt$ has SU(2)$_\stxt\times$U(1)$_\vtxt$ symmetry
and $H = H_\Ctxt+H_\vtxt+H_\Ztxt$ has U(1)$_\stxt\times$U(1)$_\vtxt$ symmetry.}
\begin{tabular}{ l || c |c || c |c }
\hline
& Symmetry of $H_\Ctxt+H_\vtxt$ & generators & Symmetry of $H$ & generators\\
\hline
$g_\perp=0$ & SU(2)$_\stxt^K\times$SU(2)$_\stxt^{K'}\times$U(1)$_\vtxt$ & $S_\al$, $N_\al$, $T_z$ &
U(1)$_\stxt^K\times$U(1)$_\stxt^{K'}\times$U(1)$_\vtxt$ & $S_z$, $N_z$, $T_z$ \\
\hline
$g_\perp=g_z$ & SU(2)$_\stxt\times$SU(2)$_\vtxt$ & $S_\al$, $T_\al$ &
U(1)$_\stxt\times$SU(2)$_\vtxt$ & $S_z$, $T_\al$\\
\hline
$g_\perp+g_z=0$ & SO(5) & $S_\al$, $T_z$, $\Pi_\al^x$, $\Pi_\al^y$ &
U(1)$_\stxt\times$SU(2) & $S_z$, $T_z$, $\Pi_z^x$, $\Pi_z^y$\\
\hline
\end{tabular}
\label{table1}
\end{table*}

\section{Exact diagonalization}
We have performed exact diagonalization (ED) studies for the Hamiltonian specified in Eq.~(\ref{Hamiltonian})
acting in a $\nu=0$ torus-geometry Hilbert space with up to $N_\phi=8$ orbitals per flavor.
When only Coulomb interactions are included, we verify that the ground state is a single Slater determinant
with two occupied and two empty flavors~\cite{YDM}.
The SU(4) multiplet structure of this broken-symmetry state is discussed in Appendix B.
We specify the ratio of $g_{z}$ to $g_{\perp}$ by the angle $\theta_g=\tan^{-1}(g_z/g_\perp)$
and fix the valley-dependent interaction strength  $g/l_B^2=\sqrt{g_\perp^2+g_z^2}/l_B^2$ at $0.01 e^2/(\epsilon l_B)$.
Because $g N_{\phi}/l_B^2$ is small compared to the Coulomb model charge-neutral energy gap that separates the ground state multiplet from the first excited multiplet at zero momentum, the role of the valley-dependent interactions is simply to lift
the Coulomb model degeneracy and split the corresponding SU(4) ground state multiplet.
Over the angle ranges $\theta_g\in[-\pi/4, \pi/2]$ and $\theta_g\in[5\pi/4, 7\pi/4]$
the exact ground states of $H_\Ctxt+H_\vtxt$ are single-Slater determinants,
with F and CDW order respectively.
For other values of $\theta_g$ valley-dependent interactions are non-trivial.

\begin{figure}[htbp]
\vspace{0cm}
 \includegraphics[width=1.00\columnwidth]{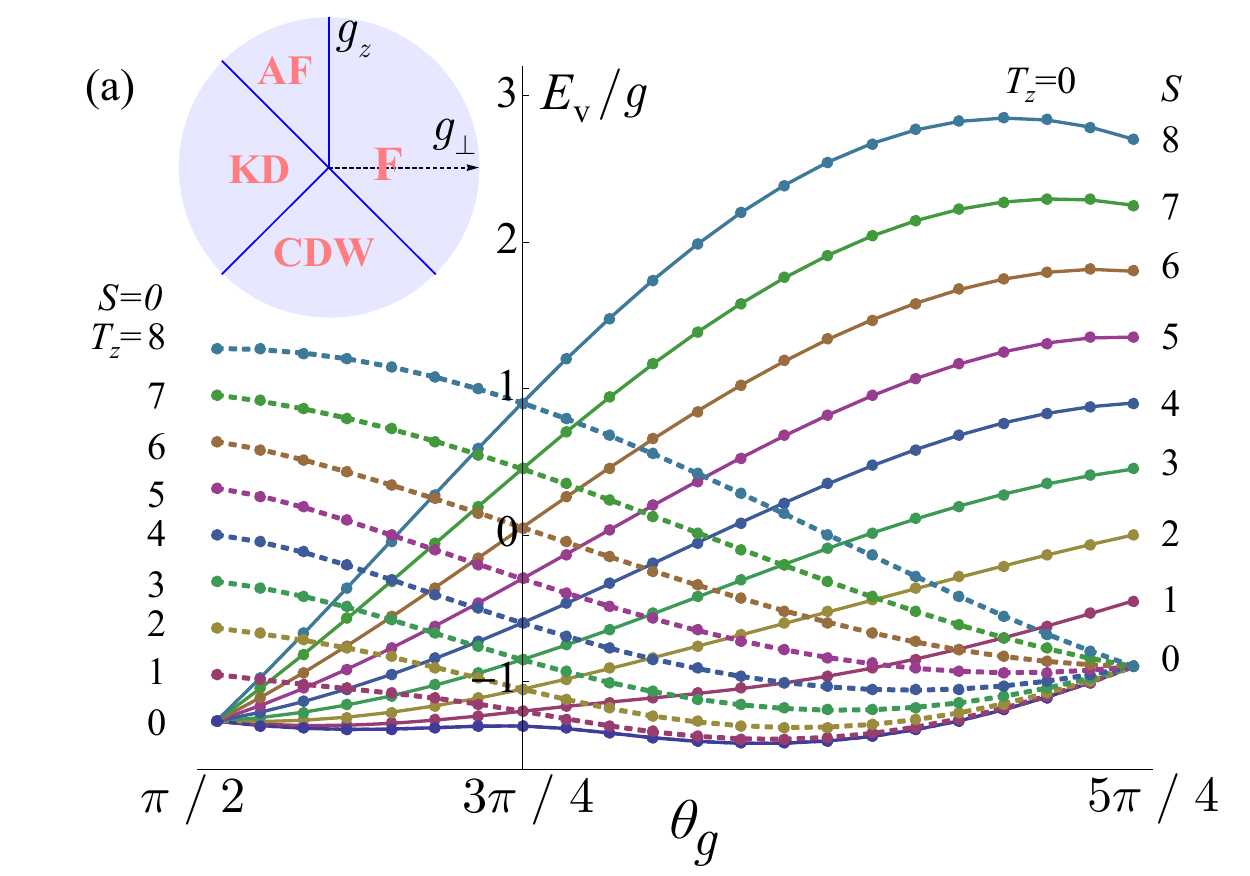}
\vspace{0cm}
\vspace{0cm}
 \includegraphics[width=1.00\columnwidth]{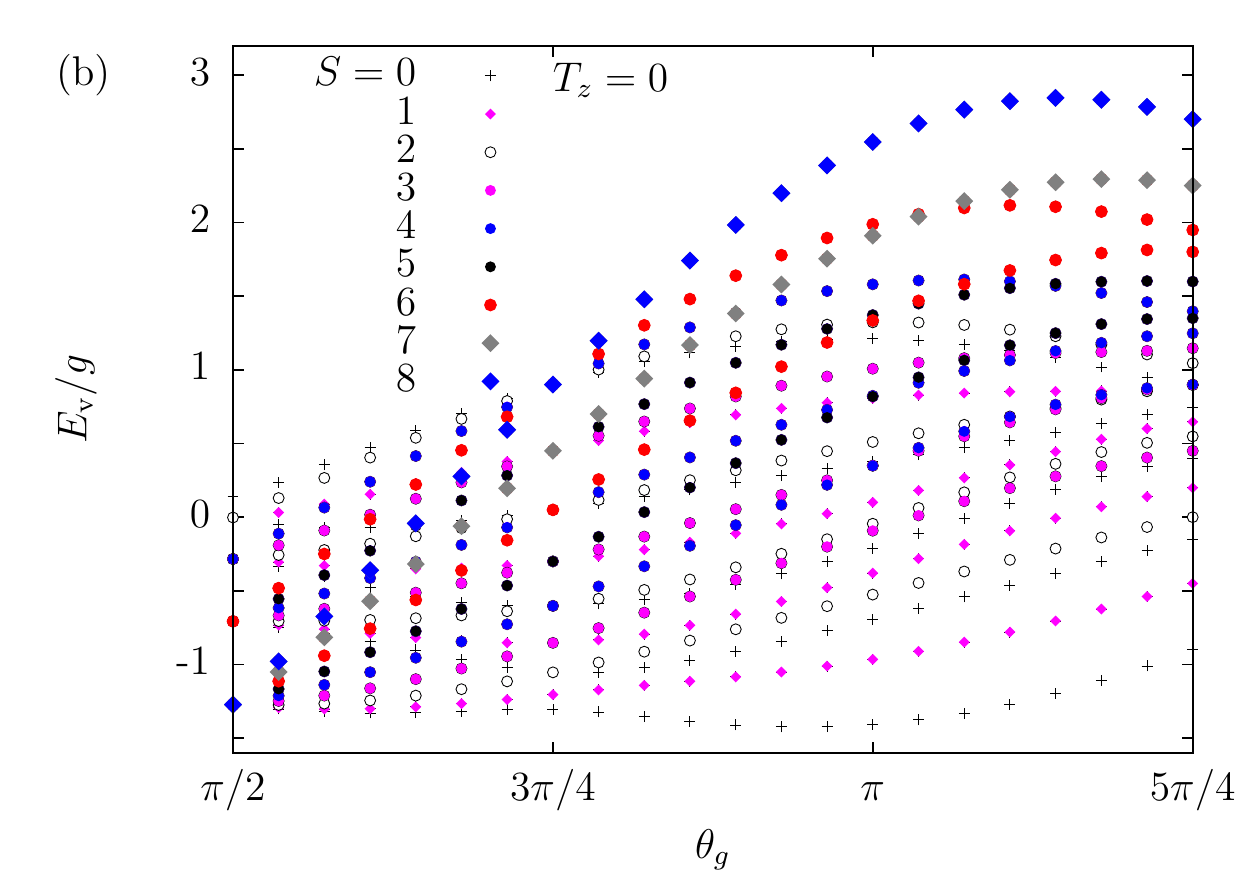}
\vspace{0cm}
\caption{Low-energy spectrum on the torus geometry for zero total momentum, filling factor $\nu=0$, and orbital Landau
level degeneracy $N_\phi=8$ as a function of $\theta_g$ in the range $[\pi/2, 5\pi/4]$.
$E_\vtxt$ is defined as the difference between the eigenvalues of $H_\Ctxt + H_\vtxt$ and the Coulomb-only
ground state energy.
All plotted eigenvalues are degenerate in the absence of $H_\vtxt$.
(a) Ground state energies in a series of $(T_z,S)$ sectors.
The solid lines show the \emph{lowest} $T_z=0$ energies for different total spin $S$ values.
Similarly, the dashed lines show the \emph{lowest} spin singlet ($S=0$) energies in
different $T_z$ sectors.  The ground state has $S=0$ and $T_z=0$ throughout the  plotted
$\theta_g$ range.  The inset shows the mean-field phase diagram
over the full $\theta_g$ range from Ref.~\onlinecite{Kharitonov_MLG}.
(b) Low-energy states in the $T_z=0$ sector for a series of total spin $S$ quantum numbers.
Note that at $\theta_g = 3 \pi/4$ states with different $S$ values are degenerate because of
the hidden SO(5) symmetry.}
 \label{spect}
\end{figure}

Fig.~\ref{spect} illustrates the $\theta_g$-dependence of the Hamiltonian spectrum for $N_e=16$ electrons in
$N=0$ Landau levels with $N_\phi=8$ over the $\theta_g \in [\pi/2, 5\pi/4]$ interval.
Fig.~\ref{spect}(a) plots ground state energies in various $(S_z,T_z)$ sectors and
demonstrates that the overall ground state has total valley polarization $T_z=0$ and total spin $S=0$ at all $\theta_g$ values
in this range.  Note that the dependence of energy on $T_z$ is suppressed as the CDW state
is approached ($\theta_{g} \to 5 \pi/4$) and that the dependence of energy on $S$ is suppressed as
the F state is approached ($\theta_{g} \to \pi/2$).
Fig.~\ref{spect}(b) illustrates how the $T_z=0$ sector of the SU(4) Coulomb ground-state multiplet
is split by $H_\vtxt$.  Since $H_\vtxt$ preserves SU(2)$_\stxt$ spin symmetry,
all energies in Fig.~\ref{spect}(a,b) occur in SU(2)$_\stxt$ multiplets.
At $\theta_g=3 \pi/4$ eigenvalues with different values of $S$ merge to form SO(5) multiplets,
each forming an irreducible representation of the SO(5) group.
(A geometric representation of the SO(5) multiplet structure is provided in Appendix B.)
All eigenstates have a definite value of the SO(5) Casimir operator~\cite{Hamermesh}
$\Gamma^2=S^2+T_z^2+\Pi^2 = l(l+3)$, with integer $l=0,1...N_\phi$.
The low-energy spectrum at $\theta_g = 3 \pi/4$ is accurately fit by the following equation~:
\beq
H_\vtxt^{\text{eff}}(\theta_g=\frac{3\pi}{4})=u_z\Big(\frac{2\Gamma^2}{N_\phi+1}-\frac{N_\phi(N_\phi+5)}{N_\phi+1}\Big),
\label{eff1}
\eeq
implying that the ground state, $|G(3\pi/4)\rangle$, is a SO(5) singlet with $\Gamma^2=0$.
It follows that the 5D order parameter vector $(T_{x, y}, N_{x, y, z})$ is maximally polarized~:
\beq
\langle T_x^2+T_y^2 +N^2\rangle_{3\pi/4}=\langle C_4 -\Gamma^2\rangle_{3\pi/4}=\langle C_4\rangle_{3\pi/4} \approx C_4^*,
\label{magvec}
\eeq
where $\langle\cdots \rangle_{3\pi/4}$ denotes expectation values in the ground state $|G(3\pi/4)\rangle$
and $C_4^*=N_\phi (N_\phi+4)$ is the value of the SU(4) Casimir operator $C_4$ in the
Coulomb model SU(4) multiplet.
The approximation leading to $C_4^*$ in Eq.~(\ref{magvec}) is validated by numerical calculation,
and also follows from the argument that
$|G(3\pi/4)\rangle$ is adiabatically connected to a state in the SU(4) multiplet.
Because $|G(3\pi/4)\rangle$ does not break SO(5) symmetry,
$\langle N_\alpha^2\rangle_{3\pi/4}=\langle T_\beta^2\rangle_{3\pi/4} \approx C_4^*/5$
with $\alpha=x, y, z$ and $\beta=x, y$.

Eq.~(\ref{eff1}) predicts that
in the thermodynamic limit $N_\phi\rightarrow \infty$, small $l$ multiplets
will approach degeneracy.
By making an analogy with the quantum rotor model, we
can see that this property signals spontaneous SO(5) symmetry breaking.
The energy in Eq.~(\ref{eff1}) can be interpreted as the kinetic energy of a generalized rotor model
in the 5D $(T_{x, y}, N_{x, y, z})$ space with the SO(5) generators playing the role of angular momenta.
In the thermodynamic limit $N_\phi\rightarrow \infty$, the moments of inertia of the rotors diverge
and it can be stuck in a spontaneously chosen direction, resulting in symmetry breaking.
The absence of ground state level crossings along the $\theta_g=3 \pi/4$ line
in Fig.~\ref{spect} indicates that the crossover between AF and KD states is smooth in finite size systems.
However, the level crossings between the low-lying excited states in Fig.~\ref{spect} signals a first order phase transition
in the thermodynamic limit.

\section{Low-energy effective theory and Collective Modes}\label{Low-energy}
Following Refs.\cite{Kharitonov_MLG,QHB,Burkov}, we can derive a
low-energy effective field theory for $\nu=0$ quantum Hall states by constructing the Lagrangian,
\beq
L =\langle\psi|i\partial_t-H|\psi\rangle=\int\frac{\dt^2 \rb}{2 \pi l_B^2} \; \big[ \Bc-\Hc \big] ,
\label{lag}
\eeq
where $|\psi\rangle$ is a Slater-determinant state in which two orthogonal
occupied spinors $\chi_{1, 2}$ are allowed to vary slowly in space and time.
The Lagrangian density $\Lc =\Bc-\Hc$ has kinetic Berry phase ($\Bc=i(\chi_1^\dg \pd_t \chi_1 + \chi_2^\dg \pd_t \chi_2)$)
and energy density $\Hc$ contributions.
As detailed in Appendix C we find that~:
\begin{widetext}
\beq
\bag
\Hc=&-u_{\perp}-2\epsilon_\Ztxt s_z + (u_z+u_{\perp}) (t_z^2 - \sum_{\al=x,y,z}s_\al^2) +2u_{\perp}\sum_{\beta=x,y}t_\beta^2 +(u_{\perp}-u_z)\sum_{\al=x,y,z}n_\al^2\\
&+l_B^2\big[\rho_z(\nabla t_z)^2+\rho_\perp\sum_{\beta=x,y}(\nabla t_\beta)^2
+\sum_{\al=x,y,z}\rho_s(\nabla s_\al)^2+\rho_n(\nabla n_\al)^2+\rho_\pi\big((\nabla \pi_\al^x)^2+(\nabla \pi_\al^y)^2\big)\big].
\label{engden}
\eag
\eeq
\end{widetext}
The stiffness coefficients $\rho_z=\rho_0-(3u_z+2u_\perp)/4$, $\rho_\perp=\rho_0-(u_z+u_\perp)/4$,
$\rho_s=\rho_0+ (u_z+2u_\perp)4$, $\rho_n=\rho_0+(u_z-2u_\perp)/4$ and $\rho_{\pi}=\rho_0-u_z/4$,
are dominated by the common Coulomb contribution $\rho_0=\sqrt{2\pi}e^2/(16\epsilon l_B)$.
It is easy to check that the energy density function $\Hc$ has the same symmetries as the Hamiltonian $H$.
The mean-field theory ground state is determined by
assuming that all fields are static and spatially uniform.
The energy competitions behind the mean-field phase diagram previously derived by Kharitonov~\cite{Kharitonov_MLG}
are transparent when Eq.~(\ref{engden}) is combined with the normalization constraint
$\sum_{\al} (t_{\al}^2 + n_{\al}^2 + s_{\al}^2+(\pi_{\al}^x)^2+(\pi_{\al}^y)^2) = 1$
(see Appendix C).
In the absence of a Zeeman field the four mean field phases are
the F state ($\sum s_{\al}^2=1$),
the AF state ($\sum n_{\al}^2=1$),
the KD state ($t_x^2+t_y^2=1$),
and the CDW state  ($t_{z}^2=1$).
The phase boundaries between these states, shown in the inset of Fig.~\ref{spect}(a),
lie along the high symmetry lines identified in Table \ref{table1}.

We now concentrate on physics near $u_z+u_\perp=0$ where
a first order phase transition occurs between KD and AF states and the
system exhibits SO(5) symmetry. The $u_z+u_\perp=0$ line in graphene is analogous to the $J_{xy} = J_{z}$ line in a $XXZ$ spin model, along which a phase transition occurs
between Ising and $XY$ ground states and the system exhibits expanded O(3) symmetry.
One physical manifestation of SO(5) symmetry along the transition line is the response to an external
Zeeman field, which induces a finite $z$ direction spin polarization $s_z$.
It follows from orthogonality constraints on the fields discussed in Appendix C that when among the
ten SO(5) generators only $s_z$ has a finite expectation value, $t_{x, y}$ and $n_z$ must vanish.
A finite Zeeman energy therefore favors the AF state over the KD state because the AF
state can distort to a canted AF with a finite $s_z$ and a N\'eel vector lying in the $xy$ plane.
A sufficiently strong Zeeman field eventually favors the F state.
Because experiments detect what appears to be a continuous phase transition as a function of Zeeman coupling strength~\cite{Pablo14},
they suggest that the ground state in the absence of Zeeman coupling lies in the AF region of the phase diagram.

Close to the
$u_z+u_\perp=0$ line, the system retains crucial SO(5) properties in the presence of a small Zeeman term.
Approximate SO(5) symmetry is revealed in the collective mode spectra of both KD and AF states.
The KD phase spontaneously breaks the valley U(1)$_\vtxt$ symmetry.
Chosing the ground state to have valley polarization $t_x$ with a spontaneous non-zero value,
we see that infinitesimal SU(4) rotations~\cite{SUN} give rise to
infinitesimal values of eight fields, $\{t_{y, z}, n_{x ,y, z}, \pi_{x, y, z}^y\}$, and leave the remaining six fields, $\{s_{x,y,z},\pi_{x,y,z}^x\}$ at zero.
The eight dynamical fields parametrize the tangent manifold of the mean-field ground state.
By evaluating the Berry phase we find that for small fluctuations the valley pseudospin fields $t_y$ and $t_{z}$ are
canonically conjugate, and that the N\'eel vector field $n_\al$ is conjugate to $\pi_\al^y$.
The valley pseudospin and N\'eel vector collective modes therefore decouple.
The valley collective mode is gapless because of the Kekul\'e state's broken
U(1) symmetry and has dispersion~:
\beq
\omega_1(\text{KD})=2 k \sqrt{\rho_\perp(u_z-u_\perp+\rho_z k^2)},
\eeq
where $k$ is wave vector and lengths are in units of $l_B$.
The three additional collective modes are kinetically coupled N\'eel-$\pi$ modes and
have energy~:
\beq
\omega_{2,3,4}(\text{KD})=2\sqrt{(| u_z + u_\perp |+\rho_n k^2)(2 |u_\perp|+\rho_\pi k^2)}.
\eeq
Note that these modes become gapless as the SO(5) symmetry line is approached and
the energy cost of N\'eel fluctuations away from the KD state vanishes, and
that the Zeeman field does not influence collective mode energies in the KD phase
because $s_z$ is not a dynamical field.
Similarly the AF state spontaneously breaks the spin SU(2)$_\stxt$ symmetry.  When the
N\'eel vector is chosen to lie along the x-axis, the dynamical fields generated by infinitesimal
SU(4) rotations are $\{s_{y, z}, n_{y, z}, t_{x, y}, \pi_x^{x, y}\}$.  Evaluating the Berry phase
we find that $s_{y}$ is conjugate to $n_{z}$ and $s_{z}$ to $n_{y}$, as in a standard
antiferromagnet.  The spin-collective modes are~:
\beq
\omega_{1,2}(\text{AF})=2k \sqrt{\rho_n(2|u_\perp|+\rho_s k^2)}.
\label{spin_wave}
\eeq
In the AF state ($t_{x}$,$\pi_x^y$)  and ($t_{y}$,$\pi_x^x$) fluctuations
form kinetically coupled conjugate pairs and give rise to the sublattice/$\pi$ collective
mode energies~:
\beq
\omega_{3,4}(\text{AF})=2\sqrt{(u_z+u_\perp+\rho_\perp k^2)(u_z-u_\perp+\rho_\pi k^2)}.
\label{pi_doublet}
\eeq
Note that all four collective modes are gapless and degenerate along the $u_z+u_\perp=0$ line.
The degeneracy arises from the SO(5) symmetry.
Appendix D describes how the collective modes in Eq.~(\ref{spin_wave}) and~(\ref{pi_doublet})
are modified by the Zeeman field.

\section{Discussion and Summary}
In ordered systems a Landau-Ginzburg or quantum effective model
which includes a single-order parameter, for example a complex pair-amplitude
order parameter for a superconductor or a magnetization direction order parameter
for a magnetic system, is often able to describe thermodynamic, fluctuation, and
response properties over wide ranges of temperature and experimentally tunable system parameters.
These theories can be powerfully predictive even when their parameters cannot
be reliably calculated from the underlying microscopic physics.
The naive effective-field-theory approach sometimes fails however.  A notable example is the case of
high-temperature superconductors in which experiments indicate that charge-density, spin-density, and
pair-amplitude order parameters have correlated quantum and thermal
fluctuations that must be treated simultaneously.  Unlike the case discussed in
the present paper in which an N=5 component  effective theory can be motivated and its parameters
estimated on the basis of microscopic physics, large-$N$ field theories\cite{SO5,Sachdev_Science_2014,Holography}
are typically constructed on the basis of hints from experimental data, for
example from observed correlations in the temperature and parameter dependence of the fluctuation amplitudes of different
observables.  In these theories, it is often difficult to be certain that all relevant fields have been identified, and
to identify constraints imposed on the fluctuations of these fields by the underlying microscopic physics.
As discussed below, the remarkably simple example of ordered states in graphene quantum Hall systems, particularly
ordered states at $\nu=0$, suggests criteria which can be tested experimentally to
validate large-N unified theories of systems with competing orders.

\begin{table}[h]
\caption{Comparison between the Kekul\'e-distortion state in graphene and the $d$-wave state in high temperature
superconductors.}
\resizebox{\columnwidth}{!}{%
\begin{tabular}{ l | c |c}
\hline
Parameter & Kekul\'e-distortion state & $d$-wave state\\
\hline
Order Parameter & ($T_x$, $T_y$) & ($\Delta_x$, $\Delta_y$)\\
\hline
U(1) generator & $T_z$ & Charge $Q$\\
\hline
External Potential & Staggered potential $\epsilon_\vtxt$ & Chemical potential $\mu$\\
\hline
\end{tabular}%
}
\label{analogies}
\end{table}

As summarized in Table~\ref{analogies}, there is a close analogy between SO(5) symmetry in the
quantum Hall effect of graphene and SO(5) symmetry in some theories
of high-$T_c$ superconductivity (HTS) ~\cite{SO5}.
The SO(5) theory of HTS theory unifies antiferromagnetism and $d$-wave superconductivity (dSC).
The analog of $d$-wave superconductivity in the graphene quantum Hall case is Kekul\'e distortion order.
The order parameters of both theories involve a sublattice degree of freedom,
the honeycomb sublattice degree-of-freedom in the case of graphene and the
sublattice degree of freedom of the magnetically ordered state in HTS SO(5) theory case.
The graphene analog of the chemical potential $\mu$ term which tunes transitions between antiferromagnetic
and $d$-wave superconducting states in the HTS case, is a sublattice-staggered potential $\epsilon_\vtxt$.
Interestingly this field is easily tunable experimentally~\cite{Weitz, Tutuc11, Velasco, Maher} in the
bilayer graphene case. SO(5) symmetry in HTS is conjectured to emerge in low-energy effective theory~\cite{SO5},
and can be exactly realized in extended Hubbard model with
artificial long-range interactions\cite{Rabello, Henley}; however, it never becomes exact for commonly used models like $t-J$ or Hubbard model.
In contrast, SO(5) symmetry and its explicit symmetry breaking naturally appear in the microscopic Hamiltonian (Eq.~(\ref{Hamiltonian}))
for the quantum Hall effect in graphene at any filling factors within $N=0$ LL.
We note that generic SO(5) symmetry without any fine-tuning parameters can appear in spin-3/2 ultracold femionic system\cite{Wu03, Hung11}.

The SO(5) symmetry in graphene is manifested by multiplet structure
in exact diagonalization spectra, and by the appearance of soft collective modes
beyond those associated with Kekul\'e or antiferromagnetic order.
In particular, the antiferromagnetic state of graphene has $\pi$-operator fluctuation
collective modes.  The observation of the analogous collective modes in the
antiferromagnetic state of high temperature superconductors would provide powerful
evidence for the applicability of an effective theory which unifies antiferromagnetism and
superconductivity only.  On the other hand their absence would likely indicate that an
effective theory of this type is not adequate over a useful range of the
tunable doping-level parameter of HTSs.  Similarly a recently proposed
alternate N=6 parameter theory\cite{Sachdev_Science_2014}
which unifies charge-density-wave and $d$-wave superconducting order,
also has implications for collective mode structure which, if verified, would
provide powerful validation.

Finally we would like to comment on the relevance of our study to the understanding of the highly insulating quantum Hall state found in graphene at neutrality. Experiments with tilted magnetic fields~\cite{Pablo14} are consistent with the view that the state at neutrality is a canted antiferromagnet. Since the transition between canted antiferromagnet and the spin polarized state is controlled solely by the ratio of the Zeeman term to the $u_\perp$ interaction strength~\cite{Kharitonov_MLG}, these very experiments serve to estimate the value of $u_\perp$, which is found to be about $u_\perp\sim -10  \epsilon_\text{Z}$~\cite{Halperin13,Inti14}. This experiment however does not serve to estimate the value of $u_z$, but simply to constrain it to satisfy $u_z \gtrsim |u_\perp|$, from the requirement that the system is in the canted Antiferromagnet phase. The determination of the value of $u_z$ relevant for monolayer graphene, and hence of its proximity to the ideal SO(5) symmetric state is therefore an open experimental problem. The presence of a weakly broken SO(5) symmetry would have important physical consequences, such as the existence of additional weakly gapped neutral collective modes as we illustrated in Section~\ref{Low-energy} and in Appendix~\ref{Zeeman}.

\section{Acknowledgments}
This work was supported by the DOE Division of Materials Sciences
and Engineering under grant DE-FG03-02ER45958, and
by the Welch foundation under grant TBF1473.
We gratefully thank Texas Advanced Computing Center(TACC)
and IDRIS-CNRS project 100383 for providing technical assistance and computer time allocations.

\appendix
\setcounter{figure}{0}
\section{Proof of SO(5) Symmetry for $g_z+g_{\perp}=0$}

Let us first briefly review how SO(5) arises naturally as a subgroup of SU(4). The fifteen generators of SU(4) can be chosen to be the Pauli matrices in spin and valley space and their direct products: $\{\sigma_\alpha,\tau_\beta,\sigma_\alpha\tau_\beta\}$.
The Clifford algebra, $\{\gamma_\mu ,\gamma_\nu\}= 2\delta_{\mu\nu}$, is realized by a subset of these generators, namely the 4x4 $\gamma$ matrices, which can be chosen as:

\beq
\bag
\ga_1=\tau_x,
\ga_2=\tau_z\sigma_x,
\ga_3&=\tau_z\sigma_y,
\ga_4=\tau_z\sigma_z,
\ga_5=\tau_y.
\eag
\eeq

SO(5) can be shown to be generated by the commutators of these $\gamma$ matrices: $[\gamma_\mu,\gamma_\nu]$. More specifically, we have the following ten generators of SO(5):
\beq
\ga_{ab}=-\frac{i}{2}[\ga_a,\ga_b]
\eeq
which can be thought of as a 5$\times$5 antisymmetric tensor~:
\begin{eqnarray}
\ga_{ab} & = & \left(
\begin{array}{ccccc}
0 &  &  &  &  \\
\tau_y \sigma_x & 0 &  &  &  \\
\tau_y \sigma_y & -\sigma_z & 0 &  &  \\
\tau_y \sigma_z & \sigma_y & -\sigma_x & 0 &  \\
-\tau_z & \tau_x \sigma_x & \tau_x \sigma_y & \tau_x \sigma_z & 0
\end{array}
\right).
\end{eqnarray}
These matrices satisfy the following commutation relations~:
\beq
[\gamma_{ab},\gamma_{cd}]=2i(\delta_{ac}\gamma_{bd}+\delta_{bd}\gamma_{ac}-\delta_{ad}\gamma_{bc}-\delta_{bc}\gamma_{ad}),
\label{SO5comm}
\eeq
\beq
[\gamma_{ab},\gamma_{c}]=2i(\delta_{ac}\gamma_{b}-\delta_{bc}\gamma_{a}).
\label{SO5vec}
\eeq
Eq.~(\ref{SO5comm}) shows that the ten independent $\gamma_{ab}$ matrices obey a set of closed commutation relations,
which is the SO(5) Lie algebra.
Additionally according to Eq.~(\ref{SO5comm}) and~(\ref{SO5vec}),
when the group is viewed as acting on $\gamma_{ab}$ and $\gamma_{a}$ by matrix conjugation, we have respectively a tensor and a vector representation of SO(5)~\cite{Georgi}.

We will now demonstrate explicitly that SO(5) is an exact symmetry of the Hamiltonian in the absence of Zeeman coupling for $g_z+g_\perp=0$. From among the fifteen generators of SU(4) identified in the main text,
the spin operator $S_\al$, the valley polarization operator $T_z$ and the $\Pi_\al^\beta$ operators
are the ten generators of the SO(5) group.
$S_\al$ and $T_z$ automatically commute with $H_\vtxt$ for any values of $g_z$ and $g_\perp$.
Thus, SO(5) will be a symmetry group if the six $\Pi_\al^\beta$ operators also commute with $H_\vtxt$.
To simplify the calculation of these commutators, we define the $\Pi$ ladder operators~:
\beq
\Pi_{\lambda'}^{\lambda}=\sum_{i}\tau_{\lambda}^i\sigma_{\lambda'}^i,
\Pi_{z}^{\lambda}=\sum_{i}\tau_{\lambda}^i\sigma_{z}^i,
\eeq
where $\lambda$ and $\lambda'$ can be $+$ or $-$.
$\tau_\pm=(\tau_x\pm i\tau_y)/2$ are ladder operators in valley space,
and the spin ladder operators $\sigma_\pm$ are similarly defined.
We work out the commutator $\big[\Pi_+^+,H_\vtxt\big]$ in detail below~:
\begin{widetext}
\beq
\bag
\big[\Pi_{+}^{+},H_{\vtxt}\big]=&\,\,
2\sum_{i\neq j}\big(-g_z\tau_z^j\tau_+^i\sigma_+^i+g_\perp\tau_+^j\tau_z^i\sigma_+^i\big)\delta(\vec{r}_i-\vec{r}_j)\\
=&\,\,
2\sum_{v,s}\sum_{p_1 p_2 p_3 p_4}
\tau_z^{vv}D_{p_1 p_2 p_3 p_4}
\big(-g_z c_{p_1 K \uparrow}^\dag c_{p_2 v s}^\dag c_{p_3 v s} c_{p_4 K' \downarrow}+
g_\perp c_{p_1 v \uparrow}^\dag c_{p_2 K s}^\dag c_{p_3 K' s} c_{p_4 v \downarrow}\big)\\
=&\,\,
2\,\, (g_z+g_\perp)\sum_{p_1 p_2 p_3 p_4}
D_{p_1 p_2 p_3 p_4}
\big(c_{p_1 K \uparrow}^\dag c_{p_2 K' \uparrow}^\dag c_{p_3 K' \uparrow} c_{p_4 K' \downarrow}
+c_{p_1 K \uparrow}^\dag c_{p_2 K \downarrow}^\dag c_{p_3 K' \downarrow} c_{p_4 K \downarrow}
\big).
\label{commutator}
\eag
\eeq
\end{widetext}
The second line of Eq.~(\ref{commutator}) is the Landau gauge second quantized form of the first line.
$c_{pvs}^\dag$ ($c_{pvs}$) is an electron creation (annihilation) operator,
$p$ denotes the orbital index within the $N=0$ Landau level, $v=K, K'$ labels valley,
and $s=\uparrow, \downarrow$ labels spin.
$D_{p_1 p_2 p_3 p_4}$ is the orbital two-particle matrix element for the $\delta$ function interaction~:
\begin{widetext}
\beq
\bag
D_{p_1 p_2 p_3 p_4}=&\int\int d\vec{r}_1 d\vec{r}_2 \,
\phi_{p_1}^*(\vec{r}_1)\phi_{p_2}^*(\vec{r}_2)
\delta(\vec{r}_1-\vec{r}_2)
\phi_{p_3}(\vec{r}_2)\phi_{p_4}(\vec{r}_1)\\
=&\int d\vec{r} \;
\phi_{p_1}^*(\vec{r})\phi_{p_2}^*(\vec{r})
\phi_{p_3}(\vec{r})\phi_{p_4}(\vec{r}),
\eag
\eeq
\end{widetext}
where $\phi_{p}(\vec{r})$ is the wave function for orbital $p$.
In the simplification leading to the last line of Eq.~(\ref{commutator}),
we used (1) fermion anticommutation relations,
and (2) the identity $D_{p_1 p_2 p_3 p_4}=D_{p_1 p_2 p_4 p_3}$,
which is a special property of $\delta$ function interaction.
Eq.~(\ref{commutator}) shows that $\big[\Pi_+^+ ,H_\vtxt\big]=0$ at $g_z+g_\perp=0$.
In a similar fashion, it can be shown that the other $\Pi$ operators also commute with $H_\vtxt$ at $g_z+g_\perp=0$.
Thus, $H_\vtxt$  has exact SO(5) symmetry for $g_z+g_\perp=0$ independent of filling factors.
The symmetry follows from the short-range nature of the valley-symmetry breaking interaction
combined with the Pauli exclusion principle for electrons.
Note that in Eq.(\ref{commutator}), we did not make use of the explicit form of the wave function $\phi_{p}(\vec{r})$.
The same Hamiltonian in Eq.(\ref{Hamiltonian}) has also been used to
describe physics in $N=0$ LL of bilayer graphene(BLG)\cite{Kharitonov_BLG,Kharitonov_edge, Kharitonov_AFM}.
There is a similar equivalence among valley, sublattice and layer degrees of freedom within $N=0$ LL in BLG.
The main difference is that $N=0$ LL in BLG contains both $n=0$ and $n=1$ magnetic oscillator states.
Since the SO(5) symmetry identified for Hamiltonian in Eq.(\ref{Hamiltonian})
is independent of single-particle wave function basis, it can also be applied to the case of BLG.

\section{Exact Diagonalization Results}

Our ED results for finite-size systems with up to 16 electrons verify that
the ground state at $\nu=0$ for Coulomb interactions only ($H=H_\Ctxt$)
is given exactly by mean field theory.
The ground state wave functions at  $\nu=0$ are
single Slater determinants with filled Landau levels for two of four flavors.
This property is a generalization of simple, quantum Hall ferromagnetism, the
occurrence of a spontaneously spin-polarized states at odd filling factors when the
spin degree-of-freedom is added to the physics of a parabolic band system Landau levels.
We have used periodic boundary conditions and classified many-body states by their magnetic translation symmetries~\cite{Haldane}.
In graphene the $\nu=0$ ground states occur at zero momentum and
form an irreducible representation of SU(4).

\begin{figure}[t]
\vspace{0cm}
 \includegraphics[width=1\columnwidth]{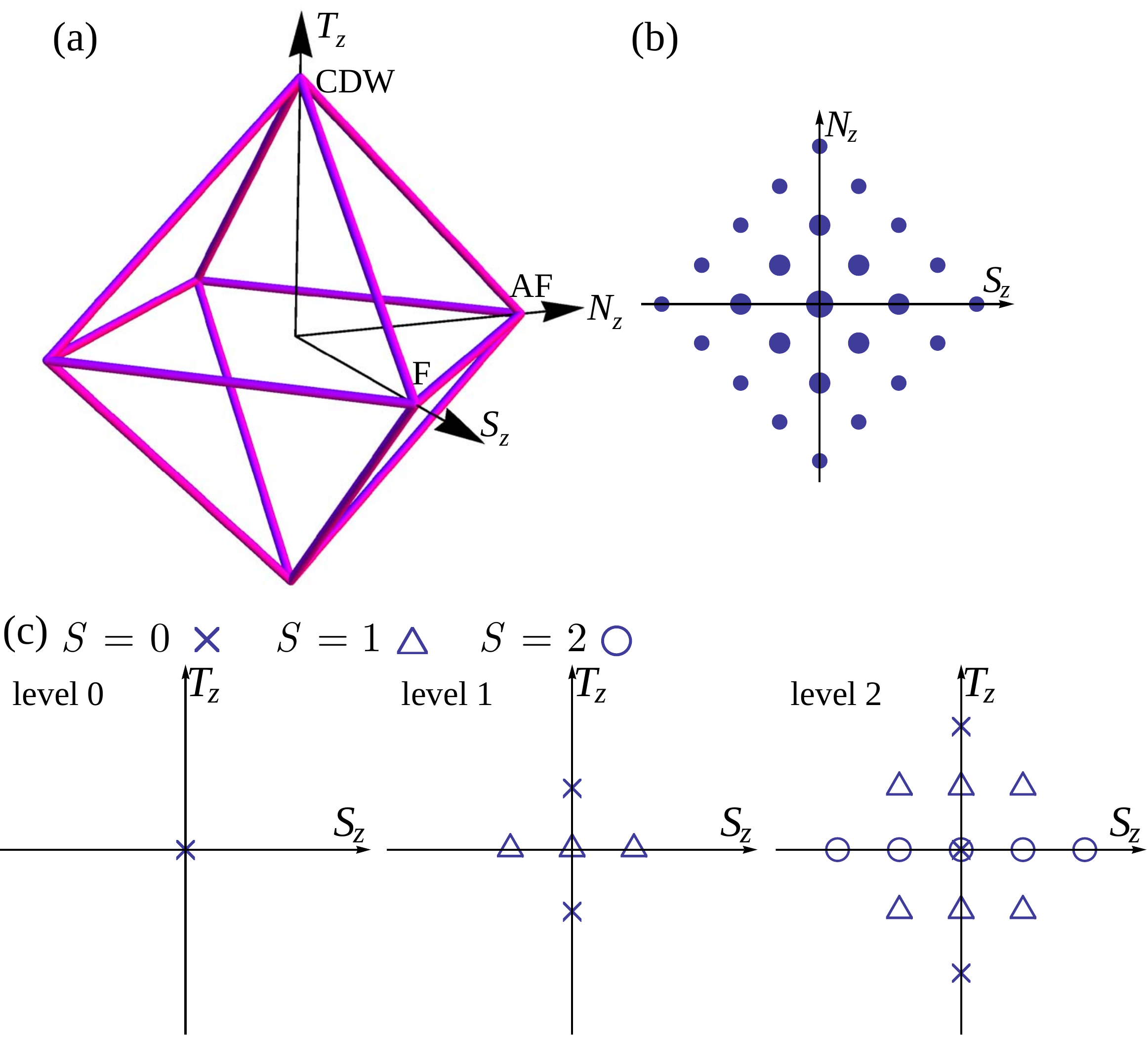}
\vspace{0cm}
\renewcommand{\figurename}{Fig.~S}
 \caption{Geometric representation of SU(4) multiplet structures.
(a) The octahedron in ($S_z, N_z, T_z$) space represents the
SU(4) multiplet structure of Coulomb ground states at $\nu=0$.
(b) A $T_z$-constant plane in the octahedron displayed for $T_z=N_\phi-4$ reached by
applying lowering operators to the CDW state with $T_z=N_\phi$.
The size of the symbols indicates the degeneracy at each point in the ($S_z, N_z$) plane.
(c) Multiplet structures of the first three levels of SO(5).}
 \label{multiplet}
\end{figure}

The $\nu=0$ F, AF and CDW states are included in the ground state multiplet and can be expressed in the form:
\beq\label{chi12}
|\chi_{1, 2}\rangle =\prod_{p=1}^{N_\phi} c^\dag_{p\chi_1}c^\dag_{p\chi_2}|0\rangle,
\eeq
where $\chi_{1, 2}$ are the two spinors defining the state and $p$ is the index of the LL orbital.
When considered as a tensor representation of SU(4),
this formula implies that the states in this multiplet are tensors with $2N_\phi$ indices in two symmetric sets each with $N_\phi$ indices
{i.e.} they are described by the Young tableau:

\begin{center}
\begin{tikzpicture}
\draw (0,0) -- (0,1) ;
\draw (0,0) -- (5,0) ;
\draw (0,1) -- (5,1) ;
\draw (5,0) -- (5,1);
\draw (0,0.5) -- (5,0.5) ;
\draw (0.5,0) -- (0.5,1) ;
\draw (4.5,0) -- (4.5,1) ;
\draw (2.5,0) node[above]{\dots\dots\dots\dots\dots};
\draw (2.5,0.5) node[above]{\dots\dots\dots\dots\dots};
\end{tikzpicture}
\end{center}
with $N_\phi$ columns and two rows.
Fig.~S\ref{multiplet}(a) represents the SU(4) multiplet structure
geometrically in terms of an octahedron in ($S_z, N_z, T_z$) space~\cite{Pfeifer}.
The octahedral shape is understood to bound a tetrahedral lattice of points in which each point designates the states
within the multiplet with common $S_z, N_z, T_z$ quantum numbers.
Fig.~S\ref{multiplet}(b) shows a slice of this lattice with $T_z=N_\phi-4$.
F, AF and CDW states are located at vertices of the octahedron,
and other orthogonal degenerate states are derived from them by applying suitable SU(4) transformations.

States in the SU(4) ground state multiplet share the same value of the SU(4) guadratic Casimir operator:
\beq
C_4=S^2+N^2+T^2+\Pi^2,
\eeq
where $S^2=\sum\limits_{\al=x,y,z}S_\al^2$, $N^2$ and $T^2$ are similarly defined,
and $\Pi^2=\sum\limits_{\al=x,y,z}(\Pi^x_\al)^2+(\Pi^y_\al)^2$.
$C_4$ takes value $N_\phi(N_\phi+4)$ for the Coulomb ground state multiplet at $\nu=0$.
Fig.~S\ref{multiplet}(b) demonstrates that there can be more than one state in the
multiplet at a given ($S_z, N_z, T_z$) point.
Hence, an additional quantum number, such as $S^2+N^2$, is needed to uniquely label a state within the SU(4)
multiplet of interest~\cite{Pfeifer}. $S^2+N^2$ is one of
the quadratic Casimir operator of the SU(2)$_\stxt^K\times$SU(2)$_\stxt^{K'}$ subgroup of SU(4).
We note that SU(2)$_\stxt^K\times$SU(2)$_\stxt^{K'}$ group has another quadratic Casimir operator $\sum\limits_{\al=x,y,z}S_\al N_\al$,
which is identical to 0 for Coulomb ground states at $\nu=0$.

\begin{figure*}[t]
\vspace{0cm}
 \includegraphics[width=1.9\columnwidth]{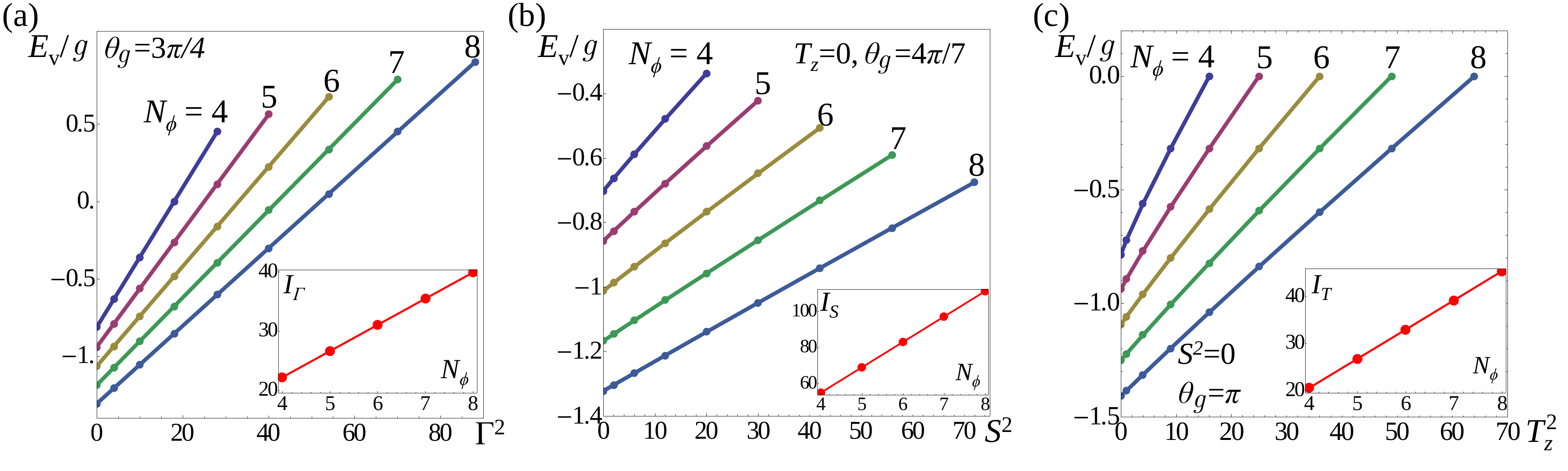}
\vspace{0cm}
\renewcommand{\figurename}{Fig.~S}
 \caption{Finite size scaling analysis.
(a) $E_\vtxt/g$ at $\theta_g=3\pi/4$ as a function of $\Gamma^2$ for $N_\phi$ ranging from 4 to 8.
(b) In a given ($T_z=0, S$) sector, the \emph{lowest} energy at $\theta_g=4\pi/7\in[\pi/2,3\pi/4]$ as a function of $S^2$.
(c) In a given ($S=0, T_z$) sector, the \emph{lowest} energy at $\theta_g=\pi$ as a function of $T_z^2$.
The inset in each figure shows the inverse of slope versus $N_{\phi}$. See text for a more detailed description.}
 \label{scaling}
\end{figure*}

SU(4) symmetry is lifted by the valley-symmetry breaking interaction $H_\vtxt$, and the octahedral multiplet is split.
At $\theta_g=3\pi/4$, SU(4) symmetry is reduced to SO(5) symmetry.
Fig.~S\ref{multiplet}(c) shows the SO(5) multiplet structure of
the three lowest energy states, which coincide with the lowest degeneracies.
Within a level, states are distinguished by $T_z$, $S_z$ and total spin $S$ quantum numbers,
and share the same value of the SO(5) Casimir operator $\Gamma^2=S^2+T_z^2+\Pi^2 = l(l+3)$, $l$ being a nonnegative integer .
We note that the same SO(5) multiplet structure has
arisen previously in numerical studies of the $t-J$ model~\cite{tjmodel}.
Interestingly, along the SO(5) line, {\it i.e.} for $\theta_g=3\pi/4$, we find
numerically that the eigenenergies are linear in $\Gamma^2$, as illustrated in Fig.~S\ref{scaling}(a).
The low-energy part of the spectrum along the SO(5) line is
accurately fit by the following equation:
\beq
H_\vtxt^{\text{eff}}(\theta_g=\frac{3\pi}{4})=u_z\Big(\frac{2\Gamma^2}{N_\phi+1}-\frac{N_\phi(N_\phi+5)}{N_\phi+1}\Big).
\label{eff1_SI}
\eeq
The ground state at $\theta_g=3\pi/4$, is an SO(5) singlet with $\Gamma^2=0$.

Away from $\theta_g=3\pi/4$, SO(5) symmetry is explicitly broken, leading to anisotropy in the 5D space. Interestingly, the spectrum can
also be fit by a linear form in the appropriate quadratic Casimir operators along other high symmetry lines.
For example, at $\theta_g=\pi/2$,  the Casimir operators of the corresponding symmetry group
SU(2)$_\stxt^K\times$SU(2)$_\stxt^{K'}\times$U(1)$_\vtxt$
are $S^2+N^2$ and $T_z^2$.
For a $T_z-$constant plane shown in Fig.~S\ref{multiplet}(b), $S^2+N^2$ takes values $f(f+2)$, with nonnegative
$f=N_\phi-|T_z|, N_\phi-|T_z|-2,\cdots$.
In analogy with the $\theta_g=3\pi/4$ case, the low-energy spectrum at $\theta_g=\pi/2$
is accurately fit by:
\beq
H_\vtxt^{\text{eff}}(\theta_g=\frac{\pi}{2})=u_z\frac{T_z^2-(S^2+N^2)+N_\phi}{N_\phi+1}.
\label{eff2}
\eeq
By interpolating between Eq.~\eqref{eff1_SI} and~\eqref{eff2}, we arrive at an
expression which describes the low-energy spectrum of the SU(4) ground state manifold over the full
$\theta_g \in (\pi/2,5\pi/4)$ interval:
\begin{widetext}
\beq
H_\vtxt^{\text{eff}}=\frac{1}{N_\phi+1}\Big(-2u_\perp\Gamma^2+(u_z+u_\perp)(T_z^2-S^2-N^2)+u_zN_\phi+u_\perp N_\phi (N_\phi+6)\Big).
\label{eff}
\eeq
\end{widetext}
Eq.~(\ref{eff}) is limited in two ways:
(1) it describes only the low-energy part of the spectrum which evolves adiabatically from the SU(4) ground state multiplet; and
(2) it is obtained by fitting numerical data at the high-symmetry points $\theta_g=\pi/2$ and $3\pi/4$.
The SO(5) symmetry-breaking states at $\theta_g=3\pi/4$ were discussed in the main text.

Eq.~\eqref{eff} makes the nature of the transition betwen AF and Kekul\'e phases at $\theta_g=3\pi/4$ clear.
As illustrated in Fig.~2(a) and discussed in the main text, the ground state throughout the entire $\theta_g \in (\pi/2,5\pi/4)$ range is singly degenerate and has $S^2=0$ and $T_z=0$.
Therefore, on the $\theta_g<3\pi/4$ side of the SO(5) line, the quantity $-(u_z+u_\perp)N^2$ in Eq.~(\ref{eff})
is an easy-plane anisotropy in the 5D space with the N\'eel vector space being the easy-plane;
N\'eel order is favored over Kekul\'e order for $\theta_g<3\pi/4$.
On the $\theta_g>3\pi/4$ side of SO(5) point,  the Kekul\'e state is favored and the
$T_{x, y}$ vectors lie in the easy-plane.
We conclude that there is a spin-flop phase transition in the 5D space across the SO(5) point. The phase transition is of first order.
Our analysis is in agreement with the mean-field prediction of a zero temperature first-order phase transition and
places it on rigorous grounds.

We will now describe how the finite size scaling demonstrates the existence of spontaneous symmetry-breaking away from the SO(5) point.
In Fig.~S\ref{scaling}(b), we plot the lowest energy at a representative angle $\theta_g=4\pi/7$ in
different $(T_z=0, S)$ sectors as a function of $S^2$
for $N_\phi$ from 4 up to 8. There is good linear relationship between the plotted energy and $S^2$.
The quantity $I_S$, defined as  the inverse of the slope, increases linearly as $N_\phi$ increases. This quantity is
a generalized moment of inertia and its divergence
indicates spontaneous SU(2)$_\stxt$ symmetry breaking in the thermodynamic limit at $\theta_g=4\pi/7$.
The reasoning is analogous as that for the SO(5) symmetry breaking at $\theta_g=3\pi/4$.
In Fig.~S\ref{scaling}(c), a similar scaling analysis
is applied to the spin singlet sector with varying $T_z$ numbers at $\theta_g=\pi$.
In this case, the analysis signals a spontaneous U(1)$_\vtxt$ symmetry breaking in the thermodynamic limit.
We remark that the finite-size scaling behavior in our system is very similar to
that in the two-dimensional antiferromagnetic Heisenberg model.
The ground state of the latter model is a spin singlet~\cite{Lieb61} in any finite size system.
However, low-lying energy levels collapse to the ground state in the thermodynamical limit,
resulting in spontaneous symmetry breaking~\cite{Bernu92,Gross89}. This set of low-lying states
is often referred to as a {\it tower of states}. 

So far, the Zeeman field has been neglected.
Since $S_z$ has been chosen as a good quantum number in our exact diagonalization calculations, the
Zeeman field simply shifts the energy of a state by an amount proportional to its $S_z$ value.
We found that the mean-field phase boundary between canted antiferromagnetic state and KD in the presence of a
Zeeman field
is in quantitative agreement with exact diagonalization results for $N_\phi=8$.

\section{Low-energy effective theory}
The continuum model Lagrangian
\beq
L =\langle\psi|i\partial_t-H|\psi\rangle=\int\frac{\dt^2 \rb}{2 \pi l_B^2} \; \big[ \Bc-\Hc \big] ,
\label{lag_SI}
\eeq
where $|\psi\rangle$ is a Slater-determinant state in which two orthogonal
occupied spinors $\chi_{1, 2}$ are allowed to vary slowly in space and time.
The Lagrangian density $\Lc =\Bc-\Hc$ has a Berry phase part~:
\beq
\Bc=i(\chi_1^\dg \pd_t \chi_1 + \chi_2^\dg \pd_t \chi_2),
\eeq
and an energy density contribution~:
\beq
\Hc=l_B^2\Ec_0(\nabla P)+\Ec_\vtxt(P)-\frac{l_B^2}{2}\Ec_\vtxt(\nabla P) +\Ec_\Ztxt(P).
\eeq
 where $P$ is the local density matrix, $P=\chi_1 \chi_1^\dg+\chi_2 \chi_2^\dg$ and
$\Ec_0(\nabla P)$ is the contribution from the SU(4) symmetric Coulomb interaction which is non-zero only when $P$ is space-dependent~:
\beq
\Ec_0(P) = \rho_0 \Tr [ \nabla P \nabla P ],
\eeq
with stiffness $\rho_0=\sqrt{2\pi}e^2/(16\epsilon l_B)$.
The next two terms are contributed by the valley-dependent interactions~:
\beq
\bag
    \Ec_\vtxt(P) &= \frac{1}{2}\sum_{\al=x,y,z} u_\al \xi_\al(P),
\eag
\eeq
where $u_{x, y} =u_\perp=g_\perp/(2\pi l_B^2)$, $u_z=g_z/(2\pi l_B^2)$,  and
$\xi_\al (P) = \Tr [\tau_\al P ] \,\Tr [\tau_\al P ] - \Tr [\tau_\al P \tau_\al P ]$.
$\Ec_\vtxt(\nabla P)$ is a gradient term,
and has a similar expression as $\Ec_\vtxt(P)$.
The last term is the Zeeman energy~:
\beq
\Ec_\Ztxt(P)=- \epsilon_\Ztxt \,\Tr [ \sigma_z P ].
\eeq
The position-dependent density matrix $P$ has the following properties~:
\beq
P^\dg = P, \quad \Tr P =2, \quad  P^2=P.
\eeq
It is convenient to reparametrize the state with a matrix $R$, where $P=\frac{1}{2}(1+R)$.
$R$ is Hermitian, traceless, and $R^2=1$.
Thus, $R$ can be expressed in terms of SU(4) generators~:
\beq
R=\sum_{a}l_a\ga_a+\sum_{a>b}l_{ab}\ga_{ab},
\eeq
where $l_a$ and $l_{ab}$ are classical real fields.
The condition $R^2=1$ gives rise to constraints on these fields.
One type is normalization constraint enforcing $\Tr[R^2]=4$~:
\beq
\sum_{a}l_a^2+\sum_{a>b}l_{ab}^2=1,
\eeq
Another type are orthogonality constraints~:
\beq
\epsilon^{abcde}l_{cd}l_e=0, \quad \epsilon^{abcde}l_{bc}l_{de}=0,
\label{orth_cons}
\eeq
where $\epsilon^{abcde}$ is the fully antisymmetric Levi-Civita symbol in five dimensions.
The orthogonality constraint is given by $\Tr[R^2\ga_{ab}]=0$ and $\Tr[R^2\ga_a]=0$.

The SO(5) theory of high-$T_c$ superconductivity~\cite{SO5} requires a similar orthogonality constraint,
which plays an essential role in predicting the phase transition between AF and dSC phases.
There, it was proposed based on a geometric interpretation of rotations in 5D~\cite{SO5},
and separately based on maximum entropy~\cite{entropy} considerations.
In our theory, the orthogonality constraint naturally appears
because of the assumption that at each LL orbital two spinors are occupied, {\it i.e.} that charge fluctuations are quenched.
To make the physical meaning of the fifteen fields $\{l_a, l_{ab}\}$ transparent, we rename them using spin and valley language~:
\beq
\bag
&l_{34, 42, 23}=s_{x, y, z},l_{1, 5} =t_{x, y},l_{15}=t_z,\\
&l_{2, 3, 4} =n_{x, y, z},\\
&l_{52, 53, 54}=\pi_{x, y, z}^x,l_{21, 31, 41}=\pi_{x, y, z}^y.
\eag
\eeq
$s_\al, t_\al$ and $n_\al$ with $\al=x, y, z$ are respectively spin, valley and N\'eel fields, and there are six $\pi$ fields.
The explicit form of the energy density $\Hc$ expressed in terms of these classical fields is given in Eq.~(\ref{engden}) of the main text.


\begin{widetext}
\section{Collective Modes in the Presence of a Zeeman Field}\label{Zeeman}
In the presence of Zeeman field, the AF is transformed to a canted antiferromagnetic (CAF) state
in which the spin-polarizations on opposite sublattices are not collinear.  In the CAF state,
the density matrix $P(\text{CAF})=\frac{1}{2}(1+\sin\theta_s\tau_z\sigma_x+\cos\theta_s\sigma_z)$
where the canting angle $\cos \theta_s=\epsilon_\Ztxt/|2u_\perp|$~\cite{Kharitonov_MLG}.
One of the spin wave mode remains gapless in the CAF state~:
\beq
\omega_1(\text{CAF})=2\sqrt{\rho_n(2|u_\perp|\sin^2\theta_s+(\rho_n\cos^2\theta_s+\rho_s\sin^2\theta_s)k^2)}k.
\eeq
This gapless mode corresponds to the rotation of N\'eel vector within the $xy$ plane.
Another spin wave mode acquires a gap~:
\beq
\omega_2(\text{CAF})=2\sqrt{(\epsilon_\Ztxt\cos\theta_s+(\rho_n\sin^2\theta_s+\rho_s\cos^2\theta_s)k^2)(2|u_\perp|+\rho_s k^2)}.
\eeq
The Zeeman field also modifies the dispersion of the sublattice/$\pi$ modes~:
\beq
\omega_{3, 4}(\text{CAF})=2\sqrt{(u_z+u_\perp+\epsilon_\Ztxt\cos\theta_s+(\rho_\perp\sin^2\theta_s+\rho_\pi\cos^2\theta_s)k^2)
(u_z-u_\perp+\rho_\pi k^2)},
\eeq
which remain gapped in
the CAF phase and become gapless at the CAF/KD phase boundary $u_z+u_\perp+\epsilon_\Ztxt\cos\theta_s=0$~\cite{Kharitonov_MLG}.

At the SO(5) point $u_z+u_\perp=0$,
the gapped spin wave mode $\omega_2(\text{CAF})$ and sublattice/$\pi$ modes $\omega_{3, 4}(\text{CAF})$ become degenerate.
The degeneracy is due to the unbroken part of the SO(5) symmetry in the presence of Zeeman field.
\end{widetext}


\begin{thebibliography}{99}

\bibitem{Kim06}
Y. Zhang, Z. Jiang, J. P. Small, M. S. Purewal, Y.-W. Tan, M. Fazlollahi, J. D. Chudow, J. A. Jaszczak, H. L. Stormer, and P. Kim,
Phys. Rev. Lett. {\bf 96}, 136806 (2006).

\bibitem{Young12}
A. F. Young, C. R. Dean, L. Wang, H. Ren, P. Cadden-Zimansky,	K. Watanabe, T. Taniguchi, J. Hone, K. L. Shepard, and P. Kim,
Nature Physics {\bf 8}, 550-556 (2012).

\bibitem{Du09}
X. Du, I. Skachko, F. Duerr, A. Luican, and E. Y. Andrei,
Nature {\bf 462}, 192-195 (2009).

\bibitem{Bolotin09}
K. I. Bolotin, F. Ghahari, M. D. Shulman, H. L. Stormer, and  P. Kim,
Nature {\bf 462}, 196-199 (2009).

\bibitem{Dean11}
C. R. Dean, A. F. Young, P. Cadden-Zimansky, L. Wang, H. Ren,	K. Watanabe, T. Taniguchi, P. Kim, J. Hone, and K. L. Shepard,
Nature Physics {\bf 7}, 693-696 (2011).

\bibitem{NM}
K. Nomura and A. H. MacDonald,
Phys. Rev. Lett. {\bf 96}, 256602 (2006).

\bibitem{YDM}
K. Yang, S. Das Sarma,  and A. H. MacDonald,
Phys. Rev. B {\bf 74}, 075423 (2006).

\bibitem{Yacoby12}
B. E. Feldman, B. Krauss, J. H. Smet, and A. Yacoby,
Science {\bf 337}, 1196 (2012).

\bibitem{Yacoby13}
B. E. Feldman, A. J. Levin, B. Krauss, D. A. Abanin, B. I. Halperin, J. H. Smet, and A. Yacoby,
Phys. Rev. Lett. {\bf 111}, 076802 (2013).

\bibitem{Pablo14}
A. F. Young, J. D. Sanchez-Yamagishi, B. Hunt, S. H. Choi, K. Watanabe, T. Taniguchi, R. C. Ashoori, and P. Jarillo-Herrero,
Nature {\bf 505}, 528-532 (2014).

\bibitem{Halperin13}
D. A. Abanin, B. E. Feldman, A. Yacoby, and B. I. Halperin,
Phys. Rev. B {\bf 88}, 115407 (2013).

\bibitem{Inti14}
I. Sodemann and A. H. MacDonald,
Phys. Rev. Lett. {\bf 112}, 126804 (2014).

\bibitem{BA}
D. M. Basko and I. L. Aleiner,
Phys. Rev. B {\bf 77}, 041409(R) (2008).

\bibitem{Fisher07}
J. Alicea and M. P. A. Fisher,
Phys. Rev. B {\bf 74}, 075422 (2006).

\bibitem{Herbut}
I. F. Herbut,
Phys. Rev. Lett. {\bf 97}, 146401 (2006).

\bibitem{Lederer07}
J.-N. Fuchs  and  P. Lederer,
Phys. Rev. Lett. {\bf 98}, 016803 (2007).

\bibitem{Jung09}
J. Jung and A. H. MacDonald,
Phys. Rev. B {\bf 80}, 235417 (2009).

\bibitem{Nomura_KD}
K. Nomura, S. Ryu, and D.-H. Lee,
Phys. Rev. Lett. {\bf 103}, 216801 (2009).

\bibitem{Mudry10}
C.-Y. Hou, C. Chamon, and C. Mudry,
Phys. Rev. B {\bf 81}, 075427 (2010).

\bibitem{Kharitonov_MLG}
M. Kharitonov,
Phys. Rev. B {\bf 85}, 155439 (2012).

\bibitem{SO5}
E. Demler, W. Hanke, and S.-C. Zhang,
Rev. Mod. Phys. {\bf 76 }, 909 (2004).

\bibitem{Aleiner07}
I. L. Aleiner,  D. E. Kharzeev, and A. M. Tsvelik,
Phys. Rev. B {\bf 76}, 195415 (2007).

\bibitem{Five_2009}
S. Ryu, C. Mudry, C.-Y. Hou, and C. Chamon,
Phys. Rev. B {\bf 80}, 205319 (2009)

\bibitem{Five_2012}
I. F. Herbut,
Phys. Rev. B {\bf 85}, 085304 (2012)

\bibitem{Kharitonov_edge}
M. Kharitonov,
Phys. Rev. B {\bf 86}, 075450 (2012).

\bibitem{Kharitonov_BLG}
M. Kharitonov,
Phys. Rev. Lett. {\bf 109}, 046803 (2012).

\bibitem{Kharitonov_AFM}
M. Kharitonov,
Phys. Rev. B {\bf 86}, 195435 (2012).

\bibitem{Hamermesh}
M. Hamermesh,
\emph{Group Theory and Its Application to Physical Problems} Ch. 8
(Dover, Reprint edition, 1989).

\bibitem{QHB}
K. Moon, H. Mori, K. Yang, S. M. Girvin, A. H. MacDonald, L. Zheng, D. Yoshioka, and S.-C. Zhang,
Phys. Rev. B {\bf 51}, 5138 (1995).

\bibitem{Burkov}
A. A. Burkov and A. H. MacDonald,
Phys. Rev. B {\bf 66}, 115320 (2002).

\bibitem{SUN}
D. P. Arovas, A. Karlhede, and D. Lilliehook,
Phys. Rev. B {\bf 59}, 13147 (1999).

\bibitem{Sachdev_Science_2014}
L. E. Hayward, D. G. Hawthorn, R. G. Melko, and S. Sachdev,
Science {\bf 343}, 1336 (2014).

\bibitem{Holography}
C. Kristjansen, R. Pourhasan, and G.W. Semenoff,
JHEP {\bf 02}, 097 (2014).

\bibitem{Weitz}
R. T. Weitz, M. T. Allen,  B. E. Feldman, J. Martin, and  A. Yacoby,
Science {\bf 330}, 812 (2010).

\bibitem{Tutuc11}
S. Kim, K. Lee, and E. Tutuc,
Phys. Rev. Lett. {\bf 107}, 016803 (2011).

\bibitem{Velasco}
J. Velasco Jr, L. Jing, W. Bao, Y. Lee, P. Kratz, V. Aji, M. Bockrath, C. N. Lau, C. Varma, R. Stillwell, D. Smirnov, F. Zhang, J. Jung, and A. H. MacDonald,
Nature Nanotechnology {\bf 7}, 156-160 (2012).

\bibitem{Maher}
P. Maher,	C. R. Dean, A. F. Young, T. Taniguchi, K. Watanabe, K. L. Shepard, J. Hone, and P. Kim,
Nature Physics {\bf 9}, 154-158 (2013).

\bibitem{Rabello}
S. Rabello, H. Kohno, E. Demler, and S.-C. Zhang,
Phys. Rev. Lett. {\bf 80}, 3586 (1998).

\bibitem{Henley}
C. L. Henley,
Phys. Rev. Lett. {\bf 80}, 3590 (1998).

\bibitem{Wu03}
C. Wu, J.-P. Hu, and S.-C. Zhang,
Phys. Rev. Lett. {\bf 91}, 186402 (2003).

\bibitem{Hung11}
H.-H. Hung, Y. Wang, and C. Wu,
Phys. Rev. B {\bf 84}, 054406 (2011).


\bibitem{Georgi}
H. Georgi,
\emph{Lie Algebras In Particle Physics: From Isospin To Unified Theories} Ch. 23 (Westview Press, 1999).

\bibitem{Haldane}
F. D. M. Haldane,
Phys. Rev. Lett. {\bf 55}, 2095 (1985).

\bibitem{Pfeifer}
W. Pfeifer,  \emph{The Lie Algebras SU(N): An Introduction} Ch. 5 (Birkh\"{a}user, 2003).

\bibitem{tjmodel}
R. Eder, W. Hanke, and S.-C. Zhang,
Phys. Rev. B {\bf 57}, 13781 (1998).

\bibitem{Lieb61}
E. Lieb, T. Schultz, and D. Mattis,
Ann. Phys.(N.Y.) {\bf 16}, 407-466 (1961).

\bibitem{Bernu92}
B. Bernu, C. Lhuillier, and L. Pierre,
Phys. Rev. Lett. {\bf 69}, 2590 (1992).

\bibitem{Gross89}
M. Gross, E. S\'anchez-Velasco, and E. Siggia,
Phys. Rev. B {\bf 39}, 2484 (1989).

\bibitem{entropy}
F. J. Wegner,
Eur. Phys. J. B {\bf 14}, 11 (2000).

\end{thebibliography}
\end{document}